\documentclass[sigconf]{acmart}

\usepackage{url}
\usepackage{graphicx}
\usepackage{booktabs}
\usepackage{balance}

\copyrightyear{2026}
\acmYear{2026}
\setcopyright{cc}
\setcctype{by-nc-nd}
\acmConference[EnCyCriS '26]{7th International Workshop on Engineering and Cybersecurity of Critical Systems }{April 12--18, 2026}{Rio de Janeiro, Brazil}
\acmBooktitle{7th International Workshop on Engineering and Cybersecurity of Critical Systems (EnCyCriS '26), April 12--18, 2026, Rio de Janeiro, Brazil}
\acmPrice{}
\acmDOI{10.1145/3786160.3788466}
\acmISBN{979-8-4007-2392-6/2026/04}

\begin{document}

\title{An Overview of Cyber Security \\ Funding for Open Source Software}

\author{Jukka Ruohonen}
\orcid{0000-0001-5147-3084}
\affiliation{\institution{University of Southern Denmark}
  \city{S{\o}nderborg}\country{Denmark}}
\email{juk@mmmi.sdu.dk}

\author{Gaurav Choudhary}
\orcid{0000-0003-3378-2945}
\affiliation{\institution{Technical University of Denmark}
  \city{Copenhagen}\country{Denmark}}
\email{gauch@dtu.dk}

\author{Adam Alami}
\orcid{0000-0003-4483-0105}
\affiliation{\institution{University of Southern Denmark}
  \city{S{\o}nderborg}\country{Denmark}}
\email{adal@mmmi.sdu.dk}

\begin{abstract}
Many open source software (OSS) projects need more human resources for
maintenance, improvements, and sometimes even their survival. These needs
allegedly apply even to vital OSS projects that can be seen as being a part of
the world's critical infrastructures. To address this resourcing problem, new
funding instruments for OSS projects have been established in recent years. The
paper examines two such funding bodies for OSS and the projects they have
funded. The focus of both funding bodies is on software security and cyber
security in general. Based on qualitative thematic analysis, the results
indicate that particularly OSS supply chains, network and cryptography
libraries, programming languages, and operating systems and their low-level
components have been funded and thus seen as critical in terms of cyber
security. In addition to the qualitative results presented, the paper makes a
contribution by connecting the research branches of critical infrastructure and
sustainability of OSS projects. A further contribution is made by connecting the
topic examined to recent cyber security regulations. Finally, an important
argument is raised that neither cyber security nor project sustainability alone
can entirely explain the rationales behind the funding decisions made by the two
funding bodies.
\end{abstract}

%

\ccsdesc[500]{Social and professional topics~Computing / technology policy}

\keywords{criticality, critical infrastructure, software security, funding,
  grants, tragedy of the commons, digital commons, incentives, regulations}

\maketitle

\section{Introduction}

The paper presents a qualitative analysis of open source software projects
funded through the Sovereign Tech Agency (STA) and a fund established by the
Open Source Security Foundation (OpenSSF). The latter is tied to the larger
Linux Foundation, while the former is operated and financed by the Germany's
central government.

For the paper's purposes, the STA is interesting because it explicitly
acknowledges that OSS is an important part of the world's critical
infrastructures---the Agency ``\textit{invests in open digital
  infrastructure}'', which refers to ``\textit{foundational technologies that
  enable the creation of other software}'', and regarding criticality, we often
do not ``\textit{notice how much our lives depend on digital infrastructure
  until it stops working}'' \cite{STA25a}. The OpenSSF's funding instrument is
based on a similar rationale; it seeks to protect ``\textit{society by
  catalyzing sustainable security improvements to the most critical open source
  software projects}''~\cite{OpenSSF24a}. Criticality is again one of the main
keywords.

To the best of the authors' knowledge, the paper is the first to examine these
two funding bodies and the OSS projects they have funded from a cyber security
perspective. In addition to the novelty of the paper's topic, the paper
contributes to two research branches. The first branch is the research on
critical infrastructure. As will be argued in the opening Section~\ref{sec:
  motivation}, this branch has often overlooked software's role in critical
infrastructures. In particular, there appears to be very limited existing work
on the relationship between open source software and critical infrastructures.

The second research branch is about the survivability or sustainability of open
source software projects~\cite{Alami24, Tulili25}. This branch has focused on a
question of what makes OSS projects survive over time. An example of a typical
explanatory factor considered in the literature is a project's ability to
attract new developers. However, regarding software quality, including software
security with it, human resources are not necessarily what matter the most;
software's growing size, complexity, and technical debt may matter
more~\cite{Alami24}. Also financial resources may well matter for the security
of OSS.

To this and other ends, recent research has also started to examine funding of
open source software projects. The paper contributes to this recent OSS funding
research and its calls for further research and better
engagement~\cite{Osborne24a, Katz25, Sharma23, Tsakpinis25}. In this regard, it
should be emphasized that the STA and OpenSSF are by no means the only funding
bodies available for OSS projects today. In addition to so-called
micro-donations and crowd-sourcing endeavors~\cite{Osborne24b, Tsakpinis25},
there are new funding initiatives for companies to support OSS
projects~\cite{FOSS25a, OSTIF25a}, philanthropic programs~\cite{CZI25a,
  NLNet25}, and funding bodies that align with civil
liberties~\cite{OTF25a}. Given the overall funding landscape, including with
respect to scientific research~\cite{Braun98, Cocos20}, these and other funding
schemes for OSS projects are strategic in their nature; they focus on some
particular problem areas. With respect to the STA and OpenSSF, the strategic
problem area is cyber security in general and software security in
particular. By implication, the paper's particular focus on the STA and OpenSSF
also connects the survivability branch with the critical infrastructure branch.

Money and cyber security have an intrinsic relationship. In fact, an argument
has often been raised that among the foremost reasons for the world's cyber
insecurity is a lack of incentives for companies, societies, and even
individuals, including software engineers~\cite{Anderson06, Halderman10}. This
lack of incentives manifests itself in observations that security is not
prioritized in the software industry because it is seen as a cost and not a
feature~\cite{Ryan23b}. Although techniques have been proposed to incentivize
and prioritize software security~\cite{Weir23}, economic factors, such as
cost-effectiveness and return-on-investment, have continued to hamper the
adoption of new security solutions for software
engineering~\cite{Ramaj24}. These obstacles are pronounced in the OSS domain
already due to the so-called tragedy of the (digital) commons~\cite{Greco04,
  OpenSSF24b}. Although this classical tragedy is a contested
concept~\cite{Frischmann19}, it serves to illustrate a common reasoning;
because OSS is openly and publicly available, there is even a less incentive to
financially invest in OSS and its security. This tragedy has been seen as a
primary reason behind the new funding initiatives~\cite{Sharma23}. The tragedy
of the (digital) commons further demonstrates that the fundamental issue is not
only about the lack of incentives but actually about misplaced
incentives. Because some OSS components are nowadays used in practically all
commercial software projects, including those addressing the world's critical
infrastructures, the misplaced incentive to not invest in the security of OSS
manifests itself in the world's overall cyber insecurity, including with respect
to digital infrastructures that are vital to the functioning of societies.

The paper's remainder is structured into six sections. The opening
Section~\ref{sec: motivation} motivates the background in more detail with
respect to the concept of critical infrastructure and its relation to
software. Given this elaboration, the subsequent Section~\ref{sec: research
  questions} presents and motivates the research questions. Then,
Section~\ref{sec: materials and methods} elaborates the empirical material and
the qualitative analysis techniques used to examine the material. Results are
presented in Section~\ref{sec: results}. A concluding discussion is presented in
Sections~\ref{sec: conclusion} and \ref{sec: further work}. With these remaining
five sections, the paper paves the way for further research on an emerging,
relevant new research area, providing also a few insights relevant to
practitioners, including OSS projects seeking funding, the two funding bodies'
staff, and policy-makers~more~generally.

\section{Motivation}\label{sec: motivation}

There is a long-standing debate regarding the concept of infrastructure and what
is recognized as critical~\cite{Eckhardt23, Katina13, Zebrowski17}. The
jurisprudence of the European Union (EU) provides a good way to briefly
illustrate typical definitions and some associated problems. Thus, the EU's now
deprecated Directive 2008/114/EC defined critical infrastructure as ``\textit{an
  asset, system or part thereof}'' that is located in any given member state and
``\textit{which is essential for the maintenance of vital societal functions,
  health, safety, security, economic or social well-being of people, and the
  disruption or destruction of which would have a significant
  impact}''. Although this definition aligns with the classical concept of an
asset, which is essential in information security~\cite{vonSolms13}, the
societal focus enlarges the scope from mere information to broader assets with
vital society-wide functions.

In this regard, the definition aligns with theoretical conceptualizations to
distinguish cyber security from information security~\cite{vonSolms13}. The
definition is also essentially sectoral~\cite{Lehto22}. However, the deprecated
directive only considered the energy and transport sectors as critical
infrastructure. In contrast, the new Directive (EU) 2022/2557 uses a similar
definition but enumerates eleven sectors as potentially critical. Although a
selection of critical sectors is left to the member states, the eleven potential
sectors highlight further conceptual expansion; among the potentially critical
sectors are energy, transport, banking and finance, healthcare, water and food
supplies, space technologies, and publication administration. In terms of
information technology, many of the Internet's core technologies are also listed
as critical, including domain name system (DNS) providers, top-level domain name
registries, cloud computing services and data centers, and content delivery
networks. The eleven critical sectors are identical to those specified in
Directive (EU) 2022/2555, which is better known as the NIS2 directive.

These EU laws have received some criticism in that too many sectors might be
covered, expertise might be lacking in some sectors, interoperability and
compliance burden may become issues, over-reporting, under-reporting, and false
positives may increase, fragmentation might occur because of sectoral
differences and the national adoptions by the member states, and sectors defined
as non-critical might receive less resources and effort~\cite{Besiekierska23,
  Ruohonen25COSE, Vandezande24, Veigurs24}. A~related critical point is that
software has received only a limited explicit attention in the laws. Although
the EU's jurisprudence recognizes information technology as a distinct critical
infrastructure of its own, it is important to emphasize that also all of the
other critical infrastructures noted are dependent on software. Neither banking
nor a central government can nowadays function without software. Nor are there
space technologies without software.

\section{Research Questions}\label{sec: research questions}

The preceding discussion motivates the first research question (RQ):
\begin{itemize}
\item{RQ.1: \textit{What kind of open source software the STA and OpenSSF have
    funded and thus considered as critical?}}
\end{itemize}

An answer to RQ.1 sheds important light on what kind of OSS is seen as critical
by a government-backed agency and a not-for-profit OSS foundation in terms of
cyber security. Indirectly, the question also aligns with existing work on how
to identify critical infrastructures~\cite{DiMauro10, Katina13}. In this regard,
it can be also remarked that the OpenSSF \cite{OpenSSF24b} has a specific
working group trying to solve the issue of how to identify critical OSS
projects. For the present purposes, an answer to RQ.1 can be reflected also from
a broader perspective; if some particular type of OSS stands out, it may be that
the EU's existing legal frameworks have overlooked some important aspects when
operating with the criticality concept.

In this regard, a further EU law can be noted: Regulation (EU) 2024/2847, which
is commonly known as the Cyber Resilience Act~(CRA). Unlike in the NIS2
directive, criticality in this regulation is not approached through sectors but
through products~\cite{Eckhardt23}. It also specifies some new cyber security
requirements for commercially supported OSS projects, including with respect to
vulnerability disclosure and associated coordination~\cite{Ruohonen25ACIG}. Open
source software foundations, which often fund OSS projects, are also mandated to
develop policies for handling vulnerabilities and fostering secure software
development. In addition, commercial vendors who integrate OSS into their
products are mandated to notify the associated OSS projects about
vulnerabilities they have discovered or been made aware of. Meeting the CRA's
essential cyber security requirements in the commercial software industry
necessitates dealing also with OSS because today many---if not most---commercial
software projects use at least some OSS components. For instance, among the
essential requirements is an obligation to ship products without known
vulnerabilities. This legal requirement applies also to all open source software
components embedded to a product.

These legal mandates can be seen to implicitly address the noted interlinking
aspect of critical infrastructures. The EU has also established new funding
schemes to support the implementation and adoption of the new cyber security
laws in the public and private sectors~\cite{Ruohonen25JKE, Veigurs24}. However,
these funding instruments seem to exclude OSS projects; hence, an answer to RQ.1
can also help to understand the point about regulators and political
decision-makers perhaps having overlooked important aspects related to critical
infrastructures and their development, maintenance, and~operation.

Regarding grants and funding, the following is worth asking:

\begin{itemize}
\item{RQ.2: \textit{What have been the monetary amounts granted and the
    durations of the projects funded by the STA and OpenSSF?}}
\end{itemize}

The money involved sheds light on the overall importance of the two funding
bodies for the OSS world in general. Regarding the funding durations, it has
been suggested to separate short-term (less than a year), medium-term (from a
year to three years), and long-term (more than three years) funding from each
other~\cite{Osborne24b}. Against this backdrop, an answer to RQ.2 can
potentially reveal insights on the cyber security issues involved in the OSS
domain. If there have been a lot of short-term funding but also some particular
long-term projects in this domain, it might be tentatively concluded that also
the security challenges vary according to a context.

The types of OSS projects funded and the money involved can reveal a lot but
obviously not everything. Particularly the wording in RQ.1 about criticality
needs sharpening. By following existing guidelines regarding funding
objectives~\cite{Osborne24b}, the point about criticality motivates the
subsequent research question:
\begin{itemize}
\item{RQ.3: \textit{What have been the rationales of the STA and OpenSSF for
    granting funding to open source software projects?}}
\end{itemize}

The question can be motivated by the extensive research on choosing OSS
components for a software project, whether commercial or open source. This
research area is closely related to the survivability and sustainability aspects
noted in the introduction. The list of factors to consider is extensive already
in terms of security alone. By and large, however, the existing research has
operated with quantified metrics that can be easily extracted from OSS
projects. To this end, an answer to RQ.3 addresses a notable limitation in
existing research because it may well be that a rationale for funding has only a
little to do with the conventional, mechanical metrics. This point entwines with
previous observations according to which developers mostly concentrate on
functionality and not security when choosing OSS components and
dependencies~\cite{Pashchenko20}. Here too the incentives noted in the
introduction are visible. In any case, an answer to RQ.3 can also be beneficial
to OSS projects and their developers in formulating competitive funding
proposals. In this regard, RQ.3 further aligns with existing research on
guidelines for writing grant applications in the domain of software
projects~\cite{Strasser22}. In addition to the types of software and the
rationales for funding decisions, it is relevant to further know what kind of
cyber security the funded projects have addressed:
\begin{itemize}
\item{RQ.4: \textit{What kind of cyber security the open source software projects funded by the STA and OpenSSF have addressed?}}
\end{itemize}

An answer to RQ.4 sheds further light on how the funding bodies have perceived
criticality in terms of cyber security. In addition, the answer helps to better
understand what kind of cyber security bottlenecks there are in the OSS domain
according to the two funding bodies. However, it may well be that the real or
perceived bottlenecks differ between the funding bodies. This point can be
extended and presented as the final research question:

\begin{itemize}
\item{RQ.5: \textit{Do the answers to RQ.1, RQ.2, RQ.3, and RQ.4 differ between
    the STA and OpenSSF?}}
\end{itemize}

An answer to the final RQ.5 serves two functions. Although the paper follows a
qualitative methodology, the first function is an overall validity check. The
second function is related to the strategic nature of the two funding bodies. If
there are notable differences, it would seem reasonable to conclude that also
the strategies of the funding bodies may differ. It could be even argued that
the strategies should differ because overlapping funding could be interpreted to
waste resources and highlight poor strategic foresight.

\section{Materials and Methods}\label{sec: materials and methods}

The data is based on funding cases that were available from the OpenSSF's
\cite{OpenSSF24a}'s and STA's \cite{STA25a} websites during the data collection
on 15 October 2025. The data includes all funding cases listed on the websites,
including those without longer funding descriptions. In total, a hundred cases
are covered; $26$ from the OpenSSF and $74$ from the STA. The periods funded
range from 2023 to 2026.

The paper's qualitative methodology is based on thematic analysis. Somewhat like
topic modeling in quantitative research, this classical qualitative method seeks
to find latent themes present in a corpus. An inductive logic was used for the
identification of themes and their refinement. By following
guidelines~\cite{Terry17, Vaismoradi13}, the thematic analysis was carried out
in five steps. The first step was about familiarizing with the qualitative data,
meaning that the funding cases were read and reread, and initial notes were
taken. Based on this familiarizing, the second step involved a generation of
initial codes for the potential themes. Then, the third and fourth steps were
about searching and identifying themes based on the codes generated, and then
collating overlapping themes due to a goal of parsimonious answers to the five
RQs specified. The fifth and final step involved defining and naming the final
themes. Although the result are summarized with concise, code-like themes
together with their occurring frequencies, the common pitfalls of mere
categorization and plain quantification~\cite{Sanders23} are addressed by
illustrating the themes with further elaborations and contemplations.

Regarding RQ.3 and RQ.4 three most important overall themes were identified for
both research questions, provided that sufficient information was
available. While these are the three most important themes in general, there is
no order of importance among the three themes themselves. This choice can be
argued to be necessary because many of the funding cases have addressed multiple
distinct topical areas akin to work packages in scientific grant applications.

Regarding coding, the initial identification of themes involved open coding,
which resulted in many themes, which were then refined and reduced by the means
of axial coding, and selective coding was finally used for collating and
sharpening the themes from the axial coding~\cite{Stol16, Williams19}. What
separates the coding done from grounded theory is that no explicit theorization
was done alongside the analysis and the themes were not taken as
latent~(cf.~\cite{Terry17}). In addition to the procedural coding logic, it is
important to emphasize that all of the paper's authors were involved in the
analysis.

It is also worth emphasizing that the coding was rather straightforward due to
the relatively short descriptions provided for the funding decisions. In
general, the STA tends to provide a little more information than the OpenSSF,
but even that information typically only amounts to a few short paragraphs. Both
funding bodies describe their decisions in unstructured formats. In addition to
the slight variance between the funding bodies, there is a variance between
individual funding decisions; some are described shortly, while others are
elaborated in detail. This variance justifies also the qualitative approach
because a quantitative approach might be biased due to the variance. The
simplicity of the final, collated themes eased the coding and its
robustness. For instance, it is rather trivial to determine whether a funded
project is about operating systems~(e.g., \cite{OpenSSF24e}), logging
libraries~\cite{STA24h}, desktop environments~\cite{STA24i}, or programming
language ecosystems (e.g., \cite{OpenSSF24d, STA24e}), and so forth.

Thus, to further elaborate the collaborative side of the thematic analysis, the
initial analysis was carried out by the paper's first author. Then, the paper's
other collaborating authors independently reviewed, case by case, the results
from the initial analysis. They were given a chance to disagree and propose
corrections, including the introduction of new themes. With respect to
disagreements and diverging themes, the first author made the final decision
about a theme by evaluating the reviews of the collaborating authors. Although
inter-rater reliability measures were omitted, this collaborative process
follows the recent suggestion to work sequentially in qualitative
coding~\cite{Diaz23}. The collaborative approach is also known as researcher
triangulation in qualitative research~\cite{Farquhar20}. Regarding inter-rater
reliability measures, the omission of these can be justified on the grounds that
only the final themes are relevant and thus reported, the prior knowledge of the
authors varied regarding the subject matter, and the funding decisions were
rather straightforward to interpret and categorize (cf.~\cite{McDonald19}; and
more generally \cite{Terry17} who argue against the inter-rater reliability
measures altogether in qualitative research).  Even though independent and
blinded coding was not done, the collaborative approach can be argued to still
improve the rigor and trustworthiness of the qualitative results.

\section{Results}\label{sec: results}

The dissemination of the qualitative results can be started by noting that $19$
out of the $100$ funding cases did not address cyber security. A good example of
non-security funding would be a grant for the GNOME desktop
environment~\cite{STA24i}. These cases are thus excluded from the answers given
to RQ.4. With this remark in mind, the results are presented by going through
the five RQs consecutively.

\subsection{Types of OSS Projects Funded (RQ.1)}

A thematic summary is presented in Table~\ref{tab: types} about the types of OSS
projects funded. As can be seen, OSS supply chains, including package managers
for installing, updating, and maintaining dependencies, have been the most
frequently funded target. The result is not surprising because the security
issues with OSS supply chains are well-recognized~\cite{Hamer25,
  Ruohonen25ARES}. These issues affect large user and deployment bases,
including in the commercial software industry due to the widespread use of OSS
components and dependencies. Supply chain security and management are also more
generally seen as being among the foremost difficult global issues in the
increasingly globalized and digitalized world~\cite{Wu24}. Therefore, it is no
wonder that the STA and OpenSSF too have recognized their criticality.

\begin{table}[th!b]
\centering
\caption{Types of OSS Funded}
\label{tab: types}
\begin{tabular}{lccc}
\toprule
Theme & Frequency \\
\hline
Supply chains and package managers & 27 \\
Programming languages and ecosystems & 21 \\
Network libraries & 12 \\
Operating systems and their components & 9 \\
Cryptographic libraries (PGP)$^1$ & 9 \\
Web (incl.~servers and frameworks) & 8 \\
Multimedia libraries & 6 \\
Scientific computing & 3 \\
Software testing frameworks & 2 \\
Logging libraries & 2 \\
Desktop environments & 1 \\
\hline
$\sum$ & 68 \\
\bottomrule
\multicolumn{2}{l}{\scriptsize{$^1$~Pretty good privacy (PGPs)}}
\end{tabular}
\end{table}

The security of OSS supply chains also aligns with the funding granted to
programming language projects because most major programming languages today
have their own ecosystems for distributing packages. A further point is that the
funding for improving OSS supply chain security aligns with the noted CRA
regulation. The reason is that particularly the OpenSSF has funded other OSS
foundations that typically govern the ecosystems and their supply chains. In the
CRA these are known as ``stewards''. Although only a lighter compliance burden
is imposed upon them by the CRA, they are still mandated to improve particularly
vulnerability disclosure and associated vulnerability coordination through
establishing policies, promoting sharing of vulnerability information, and
collaborating with European market surveillance
authorities~\cite{Ruohonen25ACIG}. To this end, it can be argued that the
funding from the STA and OpenSSF have patched a limitation in a sense that the
EU's own funding instruments have largely excluded the so-called
stewards. Further research would be needed to deduce whether the CRA was
strategically framed during its negotiations with respect to this~point.

Network libraries and protocol implementations have also been frequently funded
particularly by the STA. As could be expected, DNS and BGP implementations and
their projects are among the funding recipients. Also other networking libraries
and implementations crucial for the whole Internet are among the recipients,
including those implementing the transport layer security (TLS) and secure shell
(SSH) cryptographic protocols~\cite{OpenSSF24c, STA24c}. It is worth mentioning
also funding for a project implementing end-to-end encryption protocols as well
as four separate funding decisions for projects implementing the pretty good
privacy (PGP) cryptography protocol. Of the newer open networking protocols
related to network security, WireGuard has also received funding.

Regarding the concept of criticality, it is understandable that also operating
systems and their low-level software components have received funding. An
example would be FreeBSD~\cite{OpenSSF24e}, but the OpenSSF has also granted
funding to the Linux kernel, which is closely associated with the OpenSSF's
parent organization, the Linux Foundation. Although the data is too limited for
answering to an associated why-question, this observation is enough to remark
that the OSS funding bodies---like many but not necessarily all funding bodies
and financing instruments in general, may not be entirely immune to potential
biases and conflicts of interests.

Finally, a further point is that criticality as a concept seems to align with
the STA's \cite{STA25a} notion of foundational technologies without which the
Internet may function poorly and the creation of new software may become
difficult. The frequent funding of OSS supply chains further aligns with the
OpenSSF's \cite{OpenSSF24b} emphasis of dependencies for defining
criticality. However, there is also some room for criticism regarding the
concept of criticality and its relation to the funding decisions. For instance,
the STA \cite{STA24f} has granted $450$ thousand euros for a project seeking to
improve documentation about a JavaScript library. Even though documentation is
important, it remains debatable how critical the library is with respect to the
foundational technologies emphasized by the funding body.

\subsection{Funding Amounts (RQ.2)}

\begin{figure}[th!b]
\centering
\includegraphics[width=\linewidth, height=2.9cm]{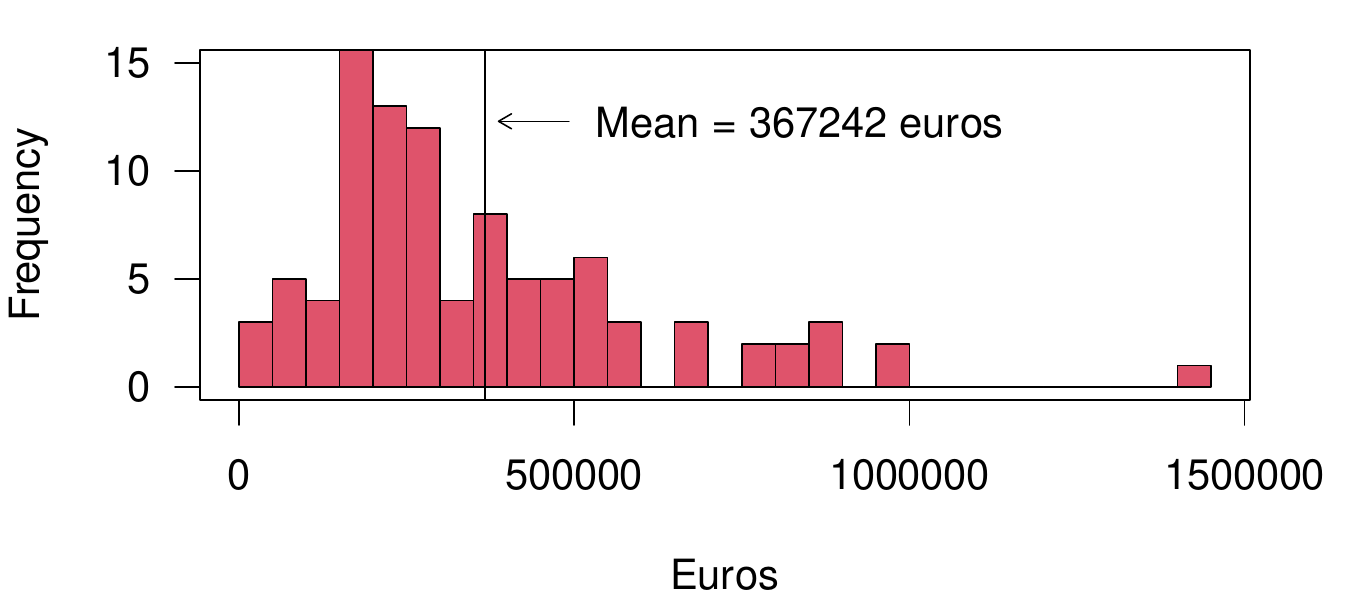}
\caption{Funding Amounts}
\label{fig: euros}
\end{figure}

The monetary amounts granted are visualized in Fig.~\ref{fig: euros}. (It can be
noted that at the time of writing the exchange rate between the EU's euro and
the United States dollar was almost even, and thus no conversion was needed.)
The average is a little below four thousand euros, but there are a few outlying
projects that have received over a million euros. Without prior work, it is
rather difficult to contemplate how large or small these amounts
are. Nevertheless, consider an imaginary example of a professional senior
software engineer earning roughly around five thousand euros a month; then, the
average funding amount would already account for about six full-time software
engineers working with a project for a year. This brief reflection allows to
argue that the funding amounts are decent enough in the OSS domain. Although the
OpenSSF does not provide data on the funding durations, most (59\%) of the STA's
funding decisions have been for two years. This observation further reinforces
the point about the interpretation of the money involved.

\subsection{Funding Rationales (RQ.3)}

The funding rationale themes are summarized in Table~\ref{tab: rationales}. To
put aside cyber security itself, which has sometimes been justified with
explicit mentions of critical infrastructure~\cite{STA24e}, also the
sustainability theme is strongly present as a rationale. Sustainability as a
theme also aligns with the rather frequent rationales of software testing and
plain old bug fixing. However, it is worthwhile to mention that some projects
have approached the sustainability conundrum through identifying and then
deprecating components and dependencies that have reached their end-of-life
software life cycle stages or became technologically risky~\cite{STA24b,
  STA24c}. In a similar vein, the point made about technical debt in the
introduction is visible in the few projects that have approached sustainability
problems through wholesale rewrites and re-implementations. Such solutions are
often interpreted as a sign of technical debt of the most severe
kind~\cite{Verdecchia21}. However, the rewrites and re-implementations have not
been only about technical debt and sustainability in general.

\begin{table}[th!b]
\centering
\caption{Funding Rationales}
\label{tab: rationales}
\begin{tabular}{lccc}
\toprule
Theme & \multicolumn{3}{c}{Frequency} \\
\cmidrule{2-4}
& STA & OpenSSF & Both \\
\hline
Cyber security & 38 & 26 & 64 \\
Sustainability & 36 & 2 & 38 \\
Testing and bug fixing & 15 & 12 & 27 \\
New features & 19 & 3 & 22 \\
Documentation and education & 12 & 3 & 15 \\
Software engineering processes & 2 & 12 & 14 \\
Infrastructure & 7 & 2 &  9 \\
Rewrites & 6 & 2 & 8 \\
Performance & 7 & 0 & 7 \\
Interoperability & 7 & 0 & 7 \\
Reliability & 5 & 0 & 5 \\
Usability & 4 & 0 & 4 \\
Auxiliary tools & 3 & 0 & 3 \\
Compliance and assurance & 1 & 0 & 1 \\
Accessibility & 1 & 0 & 1 \\
\hline
$\sum$ & 163 & 62 & 225 \\
\bottomrule
\end{tabular}
\end{table}

The reason is that some of these have involved also transformations to other
programming languages; hence, the rewrites and re-implementations also align
with memory safety. The Rust programming language is the prime example in this
regard. The language and its ecosystem have also received funding from both
bodies. From the STA \cite{STA24d}, the Rust project received funding for
implementing a software bill of materials (SBOM), while from the OpenSSF
\cite{OpenSSF24d}, it received funding for a myriad of security themes,
including hiring a dedicated security engineer, audits, documentation, advocacy,
and education, as well as new security tools and an establishment of a yet
another grant program for mentoring and education. The Rust project's own grant
program highlights potential benefits from a better coordination. The OpenSSF's
\cite{OpenSSF24d} funding case also indicates that the funding applications and
their evaluations might benefit from a more structured approach. For instance,
like in larger scientific grant applications, it could be contemplated whether
structuring the themes and their rationales into work packages would bring more
structure and coherence. A further point is that the education, outreach, and
documentation involved, and the accompanying overall theme in Table~\ref{tab:
  rationales}, frames the OSS funding question also toward academia. In this
regard, it could be argued that the security themes funded by the STA and
OpenSSF are also something that should be taught in higher education. The Rust
case further demonstrates that there may be a need in the nearby future to
update curricula with respect to programming languages.

Finally, it is important to emphasize that the funding rationales have involved
also implementing new features explicitly or implicitly related to software
security and cyber security in general. In addition, many other general software
engineering themes are present, including performance and scalability,
interoperability, reliability, usability, and software engineering processes and
practices.

\subsection{Types of Cyber Security Funded (RQ.4)}

The types of cyber security funded summarized in Table~\ref{tab: cyber security}
expectedly align closely with the earlier thematic summary in Table~\ref{tab:
  types}. In other words, supply chain security has been the most typical
thematic cyber security area funded. In this regard, it is important to further
note that some of the OSS projects funded have also addressed the design and
implementation of SBOMs. These materials for keeping track of dependencies and
other components are noteworthy for two reasons, the first being that SBOMs are
also under active research in software engineering~\cite{Bangash25}. The second
reason is that SBOMs too are present in the EU's CRA, which mandates commercial
software vendors to implement SBOMs for their products and deliver these to
market surveillance authorities upon request~\cite{Ruohonen25ACIG}. Although OSS
projects and their so-called stewards are excluded from this regulatory
requirement, the integration of OSS components into commercial software products
might partially explain also the funding decisions for SBOMs by the STA and
OpenSSF. Another, perhaps more plausible explanation might be that the funding
bodies just perceive SBOMs as important for improving supply chain security and
software security of OSS.

\begin{table}[th!b]
\centering
\caption{Types of Cyber Security Funded}
\label{tab: cyber security}
\begin{tabular}{lccc}
\toprule
Theme & \multicolumn{3}{c}{Frequency} \\
\cmidrule{2-4}
& STA & OpenSSF & Both \\
\hline
Supply chain security$^1$ & 25 & 20 & 45 \\
Cryptography & 17 & 0 & 17 \\
Audits and auditing  & 3 & 11 & 14 \\
Maintenance & 10 & 2 & 12 \\
Network security & 9 & 0 & 9 \\
Memory safety & 7 & 2 & 9 \\
SBOMs & 4 & 2 &  6 \\
Security testing & 4 & 0 & 4 \\
Security education & 1 & 2 &  3 \\
APIs$^2$ & 1 & 0 & 1 \\
\hline
$\sum$ & 81 & 39 & 120 \\
\bottomrule
\multicolumn{2}{l}{\scriptsize{$^1$~Excluding software bills of materials (SBOMs).}} \\
\multicolumn{2}{l}{\scriptsize{$^2$~Application programming interfaces (APIs).}}
\end{tabular}
\end{table}

\subsection{Divergence Between the Funds (RQ.5)}

Already the summary in Table~\ref{tab: cyber security} is enough to conclude
that there is a divergence between the funding bodies also with respect to the
types of cyber security funded. In particular, the STA has funded also many
projects addressing encryption and cryptography, network security, and security
testing, whereas the OpenSSF has been particularly active in funding of security
audits, including those done by third parties. Audits and auditing are important
to underline because these have involved also commercial cyber security
companies doing the auditing. Thus, the money involved may sometimes go to the
private sector even though the goal of funding remains the same. Furthermore,
audits and auditing align with the CRA particularly with respect to important
and critical products.

There are also some projects that have received funding from both OSS funding
bodies. The examples include FreeBSD and the Python Software Foundation, which
is maintaining the popular and large repository for distributing software
packages written in Python. The data is again too limited for far-reaching
speculations, but, nevertheless, an argument can be again raised that the two
funding bodies might benefit from joint coordination. This point can be
strengthened by noting that the data contains also some cases indicating joint
ventures between different funds~\cite{OpenSSF24c}. Against this backdrop, it
might be even reasonable to argue that the whole OSS funding landscape might
benefit from a high-level coordination body through which all funds could
collaborate and plan activities.

\section{Conclusion}\label{sec: conclusion}

The paper presented a qualitative overview of cyber security funding granted to
various OSS projects through two important funding bodies. Both bodies fund OSS
projects that are seen as being closely related to the world's critical
infrastructures. Among these software-defined infrastructures are OSS
\textit{supply chains, core computer networking libraries, servers, operating
  systems and their low-level software components, and programming languages,
  including their ecosystems}, among other things. The thematic cyber security
areas funded are related; \textit{supply chain security, cryptography,
  network security, memory safety, audits, and security education}, among other
things. Both funding bodies connect improvements at these areas to the
sustainability problems faced by many OSS projects.

However, \textit{cyber security and sustainability alone cannot fully explain
  the rationales for funding; therefore, neither is the tragedy of the (digital)
  commons sufficient as an overall theoretical explanation}. In addition to
raising this important argument, the paper made a notable contribution by
linking together the research branches addressing the sustainability of OSS
projects and critical infrastructure. Regarding the latter research branch,
software's role has often been downplayed, and the role of OSS has been almost
nonexistent in existing research. A~further contribution was made by elaborating
how the OSS cyber security funding theme examined is related to
regulations. Particularly the EU's new CRA regulation is relevant also for OSS
projects. Against this backdrop, the paper further contributed to recent efforts
to better understand the EU's new funding instruments \cite{Ruohonen25JKE}. It
seems that the two funding bodies examined have patched either intentional or
unintentional omissions in more general public sector funding in Europe and
likely also elsewhere. In a similar vein, also threat intelligence reports about
supply chain security have missed the sustainability of OSS projects as a
mitigative factor~\cite{Hamer25}.  Finally, the paper contributed to the still
nascent theorization and framing of funding for OSS.

\section{Further Work}\label{sec: further work}

The concept of criticality is the keyword behind the funding decisions. However,
as was discussed, it is not entirely clear how the two funding bodies have used
the concept in practical decision-making about funding applications. For
instance, the corresponding working group established by the OpenSSF
\cite{OpenSSF24b} has emphasized dependencies, popularity, and related concepts
in determining a project's criticality. However, dependencies and popularity may
align poorly with the STA's notions about open digital infrastructures and
foundational technologies. Consider DNS and BGP implementations as an example;
they are clearly critical for the Internet's functioning, but they may not be as
popular as some other OSS projects with large user bases. Nor do they garner a
large amount of dependencies. Thus, further research is required about the
criticality concept, including with respect to its practical use in
funding~decisions.

In addition, an important point regarding further research is the evaluation of
not only funding applications but the successes or failures of the projects
funded. In this regard, the OpenSSF \cite{OpenSSF23} has noted that their
evaluation has been ``\textit{largely anecdotal}'' and that further goals
include ``\textit{highlighting the success factors to future grant recipients
  and using them in our grant review process}''. The STA's~\cite{STA24g}
evaluation criteria for funding applications are a little more systematic;
prevalence, relevance, sustainability, public interests, activities planned and
their feasibility, and expertise of the applicants are evaluated. However, the
Agency's online materials do not say anything about whether funded open source
software projects are evaluated \textit{ex~post} after their completions. The
evaluation criteria and related points could be investigated in further research
by means of interviews, focus groups, or even surveys, given the apparent
proliferation of different funds for OSS projects.

Different evaluation criteria have been compared also in existing
research~\cite{Osborne24b}, but it remains unclear how applicable they are in
practice. Science funding provides a good example. On one hand, learning from
the funding of science projects might help with the goal of improving
evaluations. Provided that the point about potential biases and conflicts of
interests is true even to some extent, the learning might offer means to improve
objectivity in reviewing because objectivity has traditionally been highly
valued in science. For instance, it is unclear whether the STA and OpenSSF have
used external reviewers, and whether such reviewers would be seen as
preferable. On the other hand, it could be also argued that science funding
might be a poor reference because it can be bureaucratic.

In other words, lengthy and rigorous applications as well as extensive reporting
and accounting might well conflict the values and practices endorsed by OSS
communities. Analogous conflicts have been reported also in existing research
with respect to some OSS communities perceiving sponsors' and stakeholders'
interests and demands as being in a conflict with community values, norms, and
project goals~\cite{Osborne24a}. The STA and OpenSSF cases are also good
examples in a sense that the overall sustainability theme aligns only poorly
with the novelty theme usually strongly present in science funding. As has been
debated for a long time \cite{Herley17}, the other main theme, cyber security,
is also difficult to evaluate empirically.

Regardless whether science funding should or should not be used for learning, it
provides many themes for formulating relevant research questions. For instance,
a classical result has been that winning grants increases the probability of
winning more grants in the future~\cite{Bol18, Mocanu24}. It would be
interesting to know whether a similar effect applies in the OSS funding
context. The STA's \cite{STA24g} emphasis on expertise hints that a similar
effect might be present at least to some extent. The question also correlates
with the evaluation theme; it remains generally unclear what makes a competitive
funding application in the OSS context. Due to the varying domains of OSS and
the numerous distinct funds, it also remains unclear whether the guidelines
presented in existing research~\cite{Strasser22} generalize. A~further question
that needs addressing is related to the awareness among OSS communities about
the new funding opportunities available. It may be that the STA's and OpenSSF's
funding decisions have been biased not because of evaluations but because not
all critical OSS projects have been aware of the funds. Finally, as was
discussed and argued earlier, it could be evaluated whether and how academic
research might join to future funding applications. This final point can be also
used to argue that the technology transfer model~\cite{Wohlin24} may need some
altering, updating, and theorizing.

\balance
\bibliographystyle{abbrv}

\end{document}